\documentstyle[12pt,aasms4]{article}
\newcommand{\postscript}[2]
 {\setlength{\epsfxsize}{#2\hsize}
   \centerline{\epsfbox{#1}}}

\begin{document}

\title{On the Baseline Flux Determination of Microlensing Events \\
       Detectable with the Difference Image Analysis Method}

\bigskip
\bigskip
%---authors
\author{Cheongho Han}
\smallskip
\smallskip
\bigskip
\affil{Department of Astronomy \& Space Science, \\
       Chungbuk National University, Chongju, Korea 361-763 \\
       cheongho@astronomy.chungbuk.ac.kr}
\authoremail{cheongho@astronomy.chungbuk.ac.kr}

\bigskip
\bigskip

%======= ABSTRACT =================================================
\begin{abstract}
To improve photometric precision by removing blending effect, a newly 
developed technique of difference image analysis (DIA) is adopted by several 
gravitational microlensing experiment groups. However, the principal problem 
of the DIA method is that, by its nature, it has difficulties in measuring 
the baseline flux $F_0$ of a source star, causing degeneracy problem in 
determining the lensing parameters of an event. Therefore, it is often 
believed that the DIA method is not as powerful as the classical method based 
on the PSF photometry in determining the Einstein time scales $t_{\rm E}$ of 
events.

In this paper, we demonstrate that the degeneracy problem in microlensing 
events detectable from the searches by using the DIA method will not be as 
serious as it is often worried about. This is because a substantial fraction 
of events will be high amplification events for which the deviations of the 
amplification curves constructed with the wrong baseline fluxes from their 
corresponding best-fit standard amplification curves will be considerable even 
for a small amount of the fractional baseline flux deviation $\Delta F_0/F_0$. 
With a model luminosity function of source stars and under realistic 
observational conditions, we find that $\sim 30\%$ of detectable Galactic bulge
events are expected to have high amplifications and their baseline fluxes can 
be determined with uncertainties $\Delta F_0/F_0\leq 0.5$.

\end{abstract}

\vskip20mm

\centerline{resubmitted to {\it Monthly Notices of the Royal Astronomical
Society}: Sep 6, 1999}
\centerline{Preprint: CNU-A\&SS-06/99}
\clearpage

%==== Main Body ==================================================
\section{Introduction}

Detection of a large number of events is one of the big challenges in 
microlensing searches.  Classical solution to this challenge is observing 
fields with greatest density of stars such as the Galactic bulge (Alcock 
et al.\ 1997a; Udalski et al.\ 1997; Alard \& Guibert 1997) and the 
Magellanic Clouds (Alcock et al.\ 1997b, 1997c; Ansari et al.\ 1996).  
While the use of such crowded fields increases the event rate, it also 
limits the precision of the photometry due to blending (Di Stefano \& Esin 
1995; Wo\'zniak \& Paczy\'nski 1997; Han 1997; Alard 1997).  In addition, 
with the use of the classical method based on PSF photometry one can monitor 
only stars with resolved images, and thus the number of source stars is 
limited by crowding.

These problems of the classical method of microlensing experiments can be 
resolved with the newly developed technique of difference image analysis (DIA, 
Alard 1998, 1999; Alard \& Lupton 1998; Alcock et al.\ 1999a, 1999b; Melchior 
et al.\ 1998, 1999). Since the DIA method detects and measures the variation 
of source star flux by subtracting an observed image from a convolved and 
normalized reference image, one can measure light variations even for 
unresolved stars.  By using the DIA method, one can not only improve the 
photometric precision by removing the effect of blending but also increase 
the number of detected events by including unresolved stars into monitoring 
sources.  In addition, the DIA method allows one to overcome the restriction 
of conducting lensing experiments toward only resolved star fields and thus 
can extend our ability to probe extra-galactic MACHOs (Gould 1995, 1996; Han 
1996; Han \& Gould 1996; Crotts \& Tomaney 1996; Tomaney \& Crotts 1996; 
Ansari et al.\ 1997, 1999).

However, the principal problem with the DIA method in microlensing experiments 
is that, by its very nature, it has difficulties in measuring the unamplified 
flux (baseline flux, $F_0$) of a source star. This is because the observed 
light curve\footnote{Throughout this paper, we use the term `light curve' to 
designate the changes in the {\it flux} of a source star, while the term 
`amplification curve' is used to represent the changes in the {\it 
amplification} of the source star flux.} of a microlensing event obtained by 
the DIA method, $F_{\rm DIA}$, results from the combination of the true 
amplification $A_0$ and the baseline flux, i.e.\ 
$$
F_{\rm DIA}=
F-F_{\rm ref}
= F_0(A_0-1),
\eqno(1.1)
$$
where $F$ and $F_{\rm ref}$ represent the source star fluxes measured from
the image obtained during the progress of the event and the reference image, 
respectively. One significant consequence of this problem is that it produces 
degeneracy in determining the lensing parameters of the event (see \S\ 3) 
like the degeneracy problem for a blended event whose light curve results 
from the combination of $A_0$ and the blended flux. Therefore, it is often 
believed that the DIA method is not as powerful as the classical method based 
on the PSF photometry  in determining the Einstein time scale $t_{\rm E}$ of 
an event.

In this paper, we demonstrate that the degeneracy problem in microlensing 
events detectable from the searches by using the DIA method will not be as 
serious as it is often worried about. This is because a substantial fraction 
of events will be high amplification events for which the deviations of the 
amplification curves constructed with the wrong baseline fluxes from their 
corresponding best-fit standard amplification curves will be considerable even 
for a small amount of the fractional baseline flux deviation $\Delta F_0/F_0$. 
With a model luminosity function of source stars and under realistic 
observational conditions, we find that $\sim 30\%$ of detectable Galactic bulge events are expected to have high amplifications and their baseline fluxes can 
be determined with uncertainties $\Delta F_0/F_0\leq 0.5$.

\section{Mis-normalized Amplification Curves}

The standard form of the amplification curve of a gravitational microlensing 
event is related to the lensing parameters by
$$
A_0(u) = {u^2+2 \over u(u^2+4)^{1/2}};\qquad
u = \left[ \left( {t-t_{\rm max} \over t_{\rm E,0}}\right)^2 + 
\beta_0^2\right]^{1/2},
\eqno(2.1)
$$
where $u$ is the lens-source separation normalized in units of the angular
Einstein ring radius $\theta_{\rm E}$, and the lensing parameters $\beta_0$, 
$t_{\rm max}$, and $t_{\rm E,0}$ represent the impact parameter for the 
lens-source encounter, the time of maximum amplification, and the Einstein 
ring radius crossing time (Einstein time scale), respectively.  Once these 
lensing parameters are determined from the amplification curve, one can obtain 
information about the lens because the Einstein time scale is related to the 
physical parameters of the lens by
$$
t_{\rm E,0} = {r_{\rm E}\over v};\qquad
r_{\rm E}=\left( {4GM\over c^2}{D_{ol}D_{ls}\over D_{os}}\right)^{1/2},
\eqno(2.2)
$$
where $r_{\rm E}=D_{ol}\theta_{\rm E}$ is the Einstein ring radius, $v$ is 
lens-source transverse speed, $M$ is the mass of the lens, and $D_{ol}$, 
$D_{ls}$, and $D_{os}$ are the separations between the observer, lens, and 
source star.

However, if the baseline source star flux of an event is misestimated by an 
amount $\Delta F_0$, the resulting amplification curve $A$ (hereafter 
`mis-normalized' amplification curve) deviates from the true amplification 
curve $A_0$ by 
$$
A = {F_0A_0+\Delta F_0\over F_0+\Delta F_0}
={A_0+f\over 1+f}, 
\eqno(2.3)
$$
where $f=\Delta F_0/F_0$ is the fractional deviation in the determined baseline
flux.\footnote{We note that if $f$ represents the blended light fraction, i.e.\ 
$f=B/F_0$, equation (2.3) describes the observed amplification curve of a 
microlensing event affected by blended light of an amount $B$.  Therefore, the 
amplification curve of a blended event can be regarded as the mis-normalized 
amplification curve constructed with the baseline flux deviation 
$\Delta F_0=B$.  The only difference is that since the blended light should 
be positive, i.e.\ $f>0$, while the baseline flux deviation can be either 
negative or positive, blended amplification curves are always underestimated.}
If the baseline flux is overestimated (i.e.\ $f > 0$), the determined 
amplification is lower than $A_0$, and vice versa. Note that while there is 
no upper limit for $f$, it should be greater than $-1$ (i.e.\ $f > -1$).

The shape of a microlensing event amplification curve is characterized by 
its height (peak amplification) and the width (event duration), which are 
parameterized by the impact parameter and the Einstein time scale, 
respectively. Since both the height and width of the amplification curve are 
changed due to the wrong estimation of the baseline flux, the lensing 
parameters determined from the mis-normalized amplification curve will differ 
from the true values. First, the change in the peak amplification makes the 
determined impact parameter change into
$$
\beta = \left[ 2\left( 1-A_{p}^{-2}\right)^{-1/2}-2\right]^{1/2};\qquad
A_{p}={ A_{p,0}+f\over 1+f},
\eqno(2.4)
$$
where $A_{p,0}=(\beta_0^2+2)/\beta_0(\beta_0^2+4)^{1/2}$ and $A_p$ are the 
peak amplifications of the true and the 
mis-normalized amplification curves.  In addition, due to the change in the 
event duration, the determined Einstein time scale differs from the value 
$t_{\rm E,0}$ by
$$
t_{\rm E} = t_{{\rm E},0} \left( \ {\beta_{th}^2 - \beta_0^2\over
\beta_{th,0}^2 - \beta^2} \right)^{1/2},
\eqno(2.5)
$$
where $\beta_{th}$ represents the maximum allowed impact parameter (threshold 
impact parameter) for a source star to be detected by having a peak 
amplification higher than a certain threshold minimum value $A_{th}$. With 
the right choice of the baseline flux, the required minimum peak amplification 
and the corresponding maximum impact parameter are $A_{th,0}=3/\sqrt{5}$ and 
$\beta_{th,0}=1$. However, since the detectability will be determined from the 
mis-normalized amplification curve, the actually applied threshold 
amplification and the corresponding impact parameter will differ from 
$A_{th,0}$ and $\beta_{th,0}$ by
$$
A_{th}=A_{th,0}(1+f)-f,
\eqno(2.6)
$$
and 
$$
\beta_{th} = \left[ 2\left( 1-A_{th}^{-2}\right)^{-1/2}-2\right]^{1/2}
\eqno(2.7)
$$
(Han 1999).

In the upper panels of Figure 1, we present four example mis-normalized 
amplification curves $A$ (solid curves) which are expected when the baseline 
flux of the source star for a microlensing event with $\beta_0=0.5$ is 
determined with the fractional deviations of $f=\pm 0.2$ and $\pm 0.5$. By 
using equations (2.4) -- (2.7), we compute the lensing parameters of the 
standard amplification curves which best fit the individual mis-normalized 
amplification curves, and the resulting amplification curves $A_{\rm fit}$ 
are presented by dotted lines. In the lower panels, to better show the 
difference between each pair of curves $A$ and $A_{\rm fit}$, we also present 
the fractional deviations of the amplification curves $A$ from their 
corresponding best-fit standard amplification curves, i.e.\ 
$\Delta A/A_{\rm fit};\ \Delta A =A_{\rm fit}-A$. From the figure, one finds 
the following trends. First, for the same amount of 
$\left\vert f\right\vert= \left\vert \Delta F_0\right\vert/F_0$, the fractional
deviation $\Delta A/A_{\rm fit}$ is larger when the baseline flux is 
underestimated (i.e.\ $f<0$) compared to the deviation when the baseline flux 
is overestimated (i.e.\ $f>0$).  Second, although the difference between the 
two amplification curves $A$ and $A_{\rm fit}$ becomes bigger as the deviation 
$\Delta F_0/F_0$ increases, the mis-normalized amplification curves, in 
general, are well fit by standard amplification curves with different lensing 
parameters.

\section{Baseline Flux Determination for High Amplification Events} 

In previous section, we showed that since the amplification curve of a general
microlensing event obtained based on wrong estimation of the baseline flux is 
well fit by a standard amplification curve with different lensing parameters, 
making it difficult to determine $F_0$ from the shape of the observed light 
curves.  In this section, however, we show that for high amplification events 
the deviations of the mis-normalized amplification curves from their best-fit 
standard curves are considerable even for a small fractional deviation
$\Delta F_0/F_0$, and thus one can determine the baseline fluxes with small 
uncertainties.

To demonstrate this, in the upper panels of Figure 2, we present the 
mis-normalized amplification curves constructed with the same fractional 
baseline flux deviations of $f=\pm 0.2$ and $\pm 0.5$ as the cases in Figure 
1, and the corresponding best-fit standard amplification curves for a 
{\it higher} amplification event with an impact parameter of $\beta_0=0.1$. 
In the lower panels, we also present the fractional differences 
$\Delta A/A_{\rm fit}$. From the comparison of the fractional differences 
$\Delta A/A_{\rm fit}$ in Figure 1 and 2, one finds that the deviations of 
the mis-normalized amplification curves from their corresponding standard 
amplification curves are siginificantly larger for the higher amplification 
event.

To quantify how better one can determine the baseline flux with increasing  
event amplifications, we determine the uncertainty ranges of $F_0$ for 
microlensing events with various impact parameters $\beta_0$ under realistic  
observational conditions. To do this, for a given event with $\beta_0$ we first
produce a series of mis-normalized amplification curves which are constructed 
with varying values of $f$. In the next step, we obtain the best-fit standard 
amplification curves corresponding to the individual mis-normalized 
amplification curves by using the relations in equations (2.4) -- (2.7). We 
then statistically compare each pair of the amplification curves $A$ and 
$A_{\rm fit}$ by computing $\chi^2$, which are determined by
$$
\chi^2=\sum_{i=1}^{N_{\rm dat}} \left[ { A(t_i)-A_{\rm fit}(t_i)
\over p A_{\rm fit}(t_i)}\right]^2.
\eqno(3.1)
$$
For the computation of $\chi^2$, we assume that the events are observed 
$N_{\rm dat}=60$ times during $-1t_{\rm E}\leq t \leq 1t_{\rm E}$. The 
photometric uncertainty $p$ depends on the observational strategy, instrument,
and source star brightness. Therefore, we determine the photometric uncertainty
by computing the signal-to-noise ratio ($1/p=S/N$) under the assumption that 
events are observed with a mean exposure time of $t_{\rm exp}=150\ {\rm s}$ by 
using a 1-m telescope equipped with a detector that can detect 12 photons/s 
for a $I=12$ star. The detailed description about the signal-to-noise 
computation is described in \S\ 4. Once the values of $\chi^2$ as a function 
of $f$ are computed, the uncertainty of $F_0$ is determined at $1\sigma$ 
(i.e.\ $\chi^2=1$) level. We then repeat the same procedures for events with 
various values of $\beta_0$ (and thus the peak amplifications).  

In the upper of Figure 3, we present the resulting values of $\chi^2$ as a 
function of ${\rm log}(1+f)$ for example events with the source star brightness
$I=18$ and various impact parameters of $\beta_0=0.05$, 0.1, 0.2, and 0.22. In 
the lower panel, we also present the uncertainty range of $\Delta F_0/F_0$ 
(shaded region). From the figure, one finds the following trends. First, the 
uncertainty significantly decreases as the impact parameter decreases, implying
that the baseline fluxes for high amplification events can be determined with 
small uncertainties. Second, if the impact parameter becomes bigger than a 
certain critical value ($\beta_{\rm crit}$), the value of $\chi^2$ becomes less
than 1, implying that the $F_0$ cannot be determined from the shape of the 
obtained light curve. For our example events with $I=18$, this corresponds to 
$\beta_{\rm crit}=0.22$. Note that the uncertainty range $\Delta F_0/F_0$ in 
the lower panel is determined only for impact parameters yielding $\chi^2\geq1$.
Third, the upper limit of the uncertainty range is always bigger than the 
lower limit.

Knowing that $F_0$ can be determined only for high amplification events, we 
define {\it the critical impact parameter} $\beta_{\rm crit}$ as the maximum 
allowed impact parameter below which the baseline flux of the event can be 
determined with uncertainty less than $50\%$ (i.e.\ $\chi^2 \geq 1$ and 
$\Delta F_0/F_0\leq 0.5$. Then $\beta_{\rm crit}$($F_0$) represents the average
probability that the baseline flux of an event with a source star brightness 
$F_0$ can be determinded with an uncertainty less than 50\%. We compute the 
critical impact parameters for events expected to be detected towards thd 
Galactic bulge, and they are presented in the upper panel of Figure 4 as a 
function of the source star brightness in $I$ band. From the figure, one finds 
that as the source star becomes fainter, the value of $\beta_{\rm crit}$ 
decreases. This is because for a faint source event, the photometric 
uncertainty $p$ is large. Therefore, to be distinguished from standard 
amplification curves with a statistical confidence level higher than the 
required level (i.e. $\chi^2 \geq 1$), the event should be highly amplified.

\section{Fraction of High Amplification events}

In previous section, we showed that the baseline fluxes of high amplification 
events can be determined with precision.  In this section, we determine the 
fraction of high amplification events for which one can determine $F_0$ with
small uncertainties among the total microlensing events detectable by using 
the DIA method.

Under the assumption that image subtraction is perfectly conducted\footnote{
A very ingeneous image subtraction method developed by Alard \& Lupton (1998)
demostrate that it is possible to measure the variable flux to a precision 
very close to the photon noise.}, the signal measured from the subtracted image
by using the DIA technique is proportional to the variation of the source star 
flux, i.e.\ $S\propto F_0 (A_0 -1)t_{\rm exp}$. On the other hand, the noise of
the source star flux measurements comes from both the lensed source star and 
blended stars, i.e.\ $N\propto (F_0A_0 + B)^{1/2}$, where $B$ is the background
flux from blended stars in the effective seeing disk (i.e.\ the 
undistinguishable separation between images) with a size (i.e.\ diameter) 
$\Delta\theta_{\rm see}$. Then the signal-to-noise ratio of an event whose 
light variation is detected by using the DIA method is given by
$$
S/N
=F_0(A_0-1)\left( {t_{\rm exp}\over F_0 A_0+\langle B\rangle}\right)^{1/2},
\eqno(4.1)
$$
where $\langle B\rangle$ represents the mean background flux. For a high 
amplification event ($A_0\gg 1$) with a bright source star ($F_0A_0\gg \langle 
B\rangle$), the signal-to-noise ratio becomes photon limited, i.e.\  
$S/N\sim (F_0A_0 t_{\rm exp})^{1/2}$. By contrast, for a low amplification 
event with a faint source star ($F_0A_0\ll \langle B\rangle$), the noise from 
the background flux becomes important. Let us define $\beta_{\rm max}(F_0)$ as 
the maximum impact parameter within which a lensing event can be detected by 
having signal-to-noise ratios higher than a certain threshold value 
$(S/N)_{\rm th}$ during a range of time longer than a required one $\Delta t$.
Then, $\beta_{\rm max}(F_0)$ represents the average detection probability of 
an event with a source star brightness $F_0$ from the microlensing search by 
using the DIA method, and it is computed by
$$
\beta_{\rm max}= 
\cases{
\left[ u_{\rm max}^2 - (\Delta t/t_E)^2\right]^{1/2} & 
when $u_{\rm max}\geq \Delta t/t_{\rm E}$ \cr
0 & when $u_{\rm max} < \Delta t/t_{\rm E}$ \cr
}
\eqno(4.2)
$$
where $u_{\rm max}=[2(1-A_{\rm min}^{-2})^{-1/2}-2 ]^{1/2}$ represents the 
threshold lens-source separation below which the signal-to-noise ratio of an 
event becomes greater than $(S/N)_{th}$ and $A_{\rm min}$ is the amplification 
of the event then $u$=$u_{\rm max}$. The value of $A_{\rm min}$ is obtained by 
numerically solving equation (4.1) with respect to the amplification for a 
given threshold signal-to-noise ratio of $(S/N)_{\rm th}$.

In the upper panel of Figure 4, we present the maximum impact parameter 
$\beta_{\rm max}$ as a function of the source brightenss for stars in the 
Galactic bulge.  For the computation of $\beta_{\rm max}(F_0)$, we 
assume the same observational conditions described in \S\ 3. The adopted 
threshold signal-to-noise ratio of the MACHO experiment is $(S/N)_{\rm th}=10$ 
(Alcock 1999a, 1999b). In our computation, however, a higher value of 
$(S/N)_{\rm th}=15$ is adopted to account for the additional noise from the 
sky brightness and the residual flux due to imperfect image subtraction. The 
average background flux is obtained by
$$
\langle B \rangle = \int_0^{F_{\rm CL}} F_0 \Phi_0(F_0)dF_0,
\eqno(4.3)
$$
where $F_{\rm CL}$ and $\Phi_0 (F_0)$ are the crowding limit of the Galactic 
bulge field and the luminosity function of stars in the field normalized to 
the area of $\pi(\Delta\theta_{\rm see}/2)^2$.  We adopt the luminosity 
function of Holtzman et al.\ (1998) constructed from the observations of 
bulge stars by using the {\it Hubble Space Telescope} and the adopted crowding 
limit is $I=18.2$ mag. We assume that an event is detectable if signal-to-noise
ratios are higher than $(S/N)_{\rm th}$ during $\Delta t=0.2t_{\rm E}$ of the 
source star flux variation measurements. From the figure, one finds that the 
detection probabilities (i.e.\ $\beta_{\rm max}$) of events with source stars 
faint than the crowding limit, and thus unresolvable, are not negligible up to 
$\sim 3$ mag below $F_{\rm CL}$ implying that a substantial fraction of events 
detectable by using the DIA method will be faint source star events (Jeong, 
Park, \& Han 1999).

With the determined values of $\beta_{\rm max}$ as a function of source star 
brightness, we then construct the effective source star luminosity function by
$$
\Phi_{\rm eff}(F_0) = \beta_{\rm max}(F_0)\Phi_0(F_0).
\eqno(4.4)
$$
We also construct the luminosity function of source stars for high 
amplification events with measurable baseline fluxes by
$$
\Phi_{\rm high} = \eta(F_0)\Phi_0(F_0);\qquad
\eta=
\cases{
\beta_{\rm crit}/\beta_{\rm max} & 
when $\beta_{\rm crit}\leq \beta_{\rm max}$ \cr
1.0 & when $\beta_{\rm crit} > \beta_{\rm max}$ \cr
}
\eqno(4.5)
$$
Once the two luminosity functions $\Phi_{\rm eff}$ and $\Phi_{\rm high}$ are
constructed, the fraction of high amplification events out of the total number
of detactable events is computed by
$$
{\Gamma_{\rm high} \over \Gamma_{\rm tot}} = 
{\int_0^\infty \Phi_{\rm high}(F_0) dF_0
\over 
\int_0^\infty \Phi_{\rm eff}(F_0) dF_0}.
\eqno(4.6)
$$
We find that for $\sim 33\%$ of events detectable from the microlensing 
searches by using the DIA method will have high amplification events for which 
the baseline fluxes can be determined with uncertainties 
$\Delta F_0/F_0\leq 0.5$.

\section{Conclusion}

We have investigated how the lensing parameters change due to the wrong 
determination of the baseline flux of a microlensing event. We have also 
investigated the feasibility of the baseline flux determination from the shape 
of the observed light curve. The results of these investigations are as 
follows:
\begin{enumerate}
\item
The obtained amplification curve of a general microlensing event based on wrong baseline
flux is well fit by a standard amplification curve with different lensing 
parameters, implying that precise determination of $F_0$ from the shape of the 
observed light curve will be difficult.
\item
However, for a high amplification event, the mis-normalized amplification 
curve deviates from the standard form by a considerable amount even for a small
fractional deviation of baseline flux, allowing one to determine $F_0$ with a 
small uncertainty. 
\item
With a model luminosity function of Galactic bulge stars and under realistic 
observational conditions of the microlensing searches with the DIA method, we 
find that a substantial fraction ($\sim 33\%$) of microlensing events 
detectable by using the DIA method will be high amplification events, for 
which the baseline fluxes of source stars can be determined with uncertainties 
$\Delta F_0/F_0\leq 50\%$. 

\end{enumerate}

\acknowledgements
This work was supported by the grant (1999-2-113-001-5) of the Korea Science 
\& Engineering Foundation.

\clearpage

%%%% FIGURES

\postscript{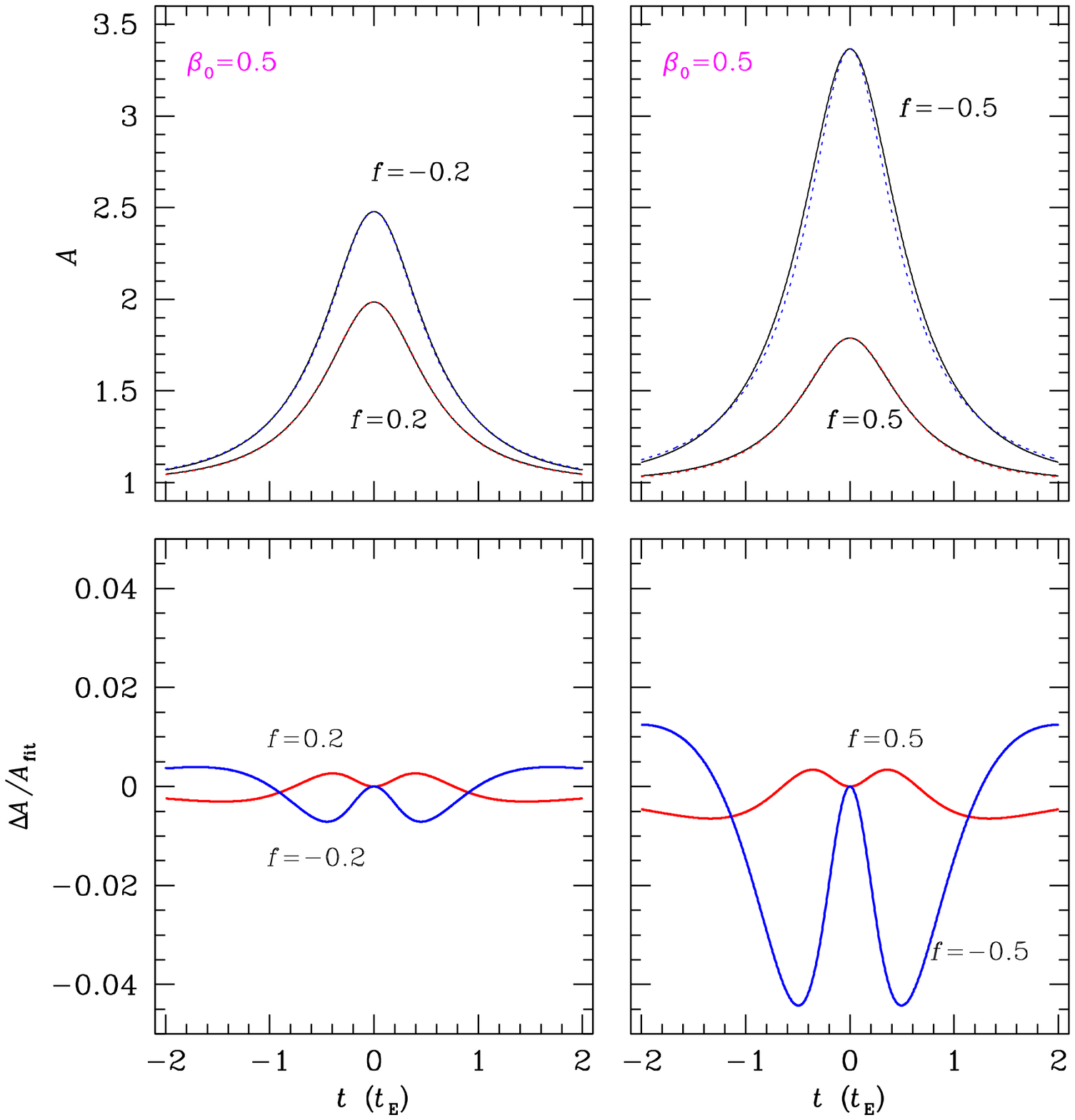}{0.95}
\noindent
{\footnotesize {\bf Figure 1:}\ 
Upper panels: 
four example mis-normalized amplification curves (solid curves) which are 
expected when the baseline flux of the source star for a microlensing event 
with $\beta_0=0.5$ is determined with the fractional deviations of $f=\pm 0.2$ 
and $\pm 0.5$.  Also presented are the standard amplification curves 
($A_{\rm fit}$, dotted curves) which best fit the individual mis-normalized 
amplification curves.
Lower panels:
the fractional deviations of the mis-normalized amplification curves 
from their corresponding best-fit standard amplification curves, i.e.\ 
$\Delta A/A_{\rm fit};\ \Delta A=A_{\rm fit}-A$.
}\clearpage

\postscript{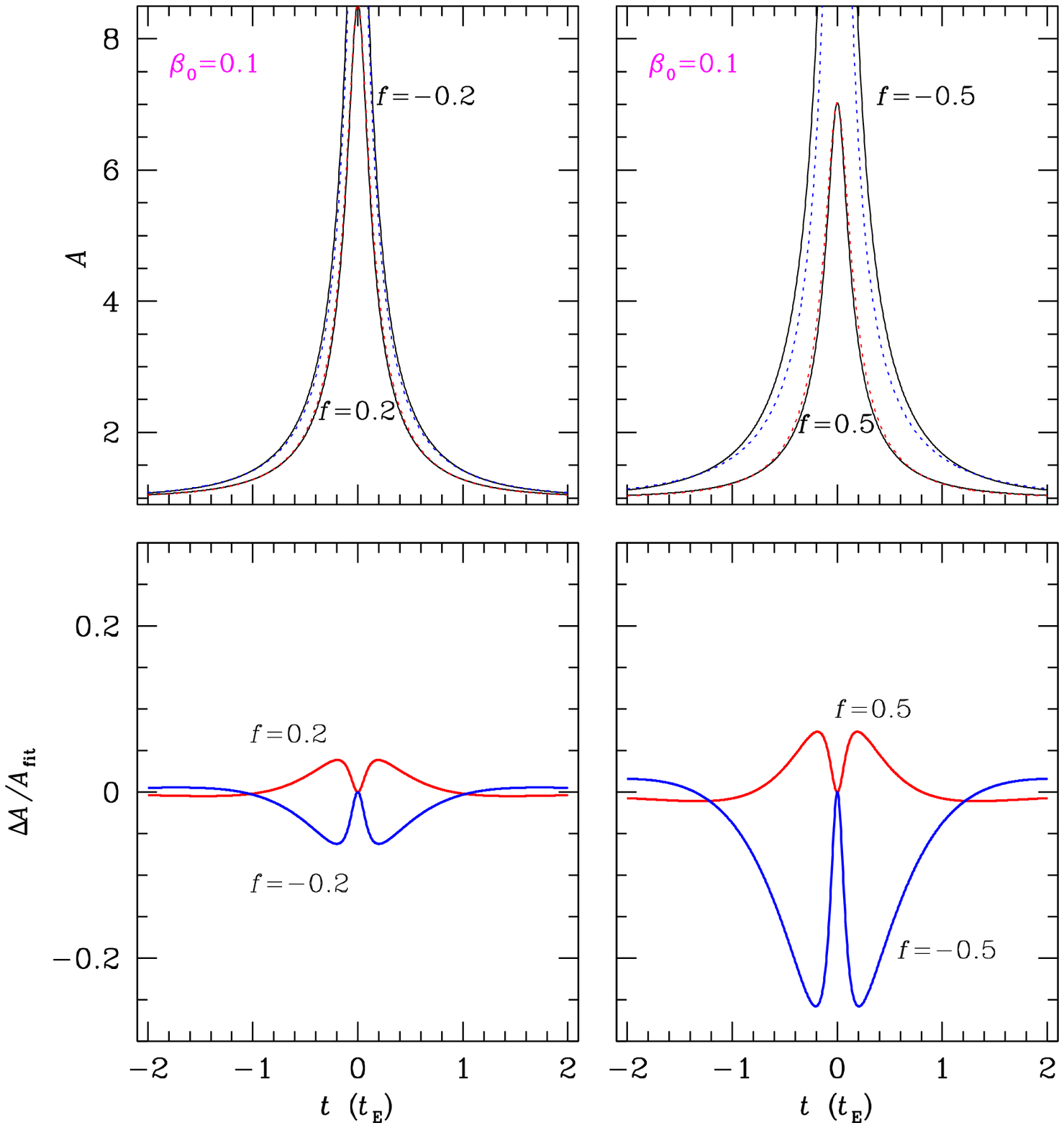}{0.95}
\noindent
{\footnotesize {\bf Figure 2:}\
Upper panel: 
the mis-normalized amplification curves with the same fractional baseline
flux deviations of $f=\pm 0.2$ and $\pm 0.5$ as the cases in Figure 1, and
the corresponding best-fit standard amplification curves for a {\it higher}
amplification event with $\beta_0=0.1$.
Lower panel:
the fractional differences between the amplification curves $A$ and 
$A_{\rm fit}$.
}\clearpage

\postscript{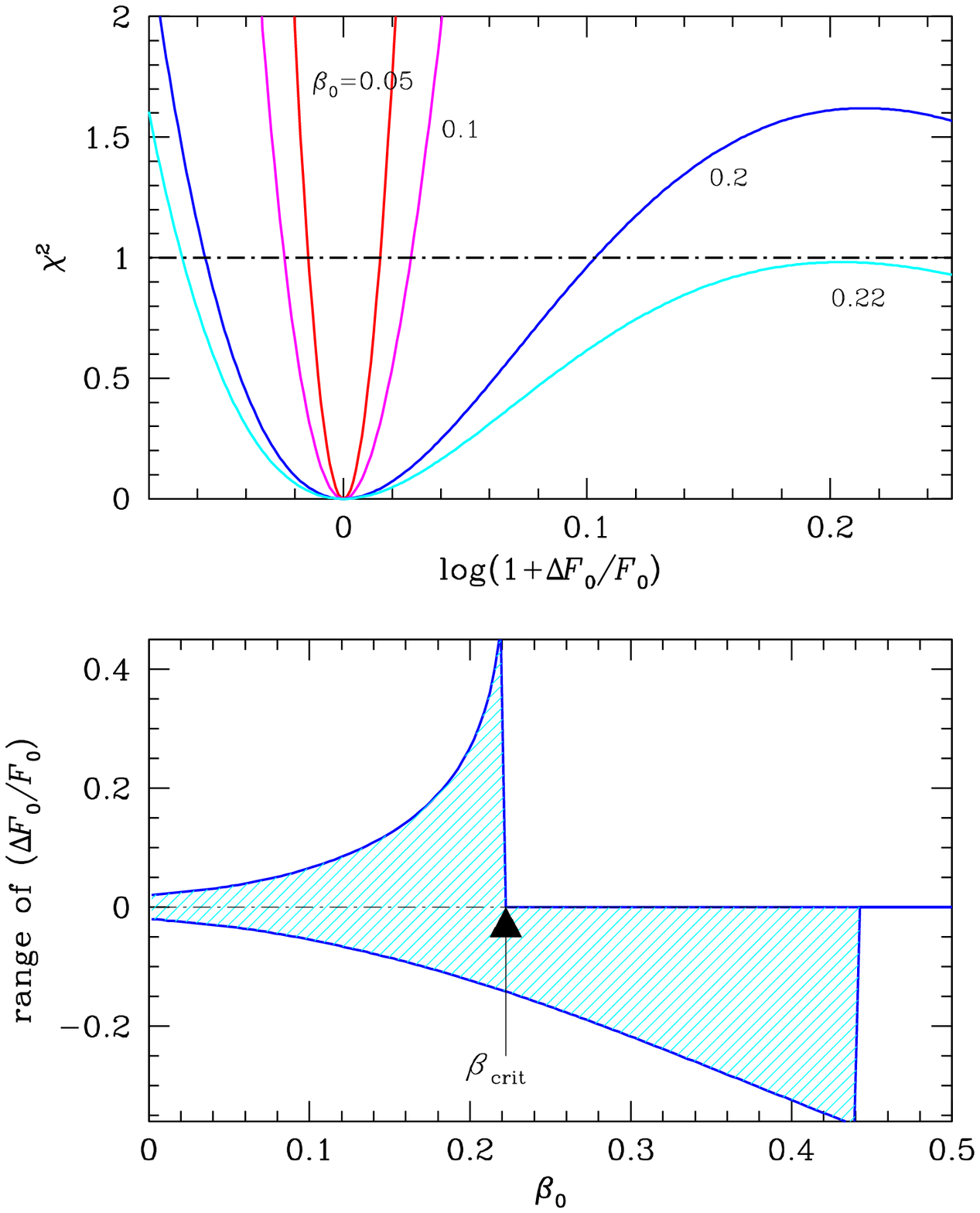}{0.95}
\noindent
{\footnotesize {\bf Figure 3:}\
Upper panel: 
the values of $\chi^2$ as a function of ${\rm log}(1+\Delta F_0/F_0)$ for 
example events with a source star brightness of $I=18$ and various impact 
parameters $\beta_0$. The value of $\chi^2$ is computed by comparing the 
mis-normalized amplification curve with a fractional baseline flux deviation 
$f=\Delta F_0/F_0$ and the corresponding best-fit standard amplification curve.
%For the computation of $\chi^2$, we assume that the event is observed 60 times 
%during $-1t_{\rm E}\leq t\leq 1t_{\rm E}$.
Lower panel: 
the uncertainty range of the baseline flux (shaded region) determined from 
the shape of the light curve of a lensing event detected by using the DIA 
technique.  The uncertainties are determined at $1\sigma$ (i.e.\ $\chi^2=1$) 
level.
}\clearpage

\postscript{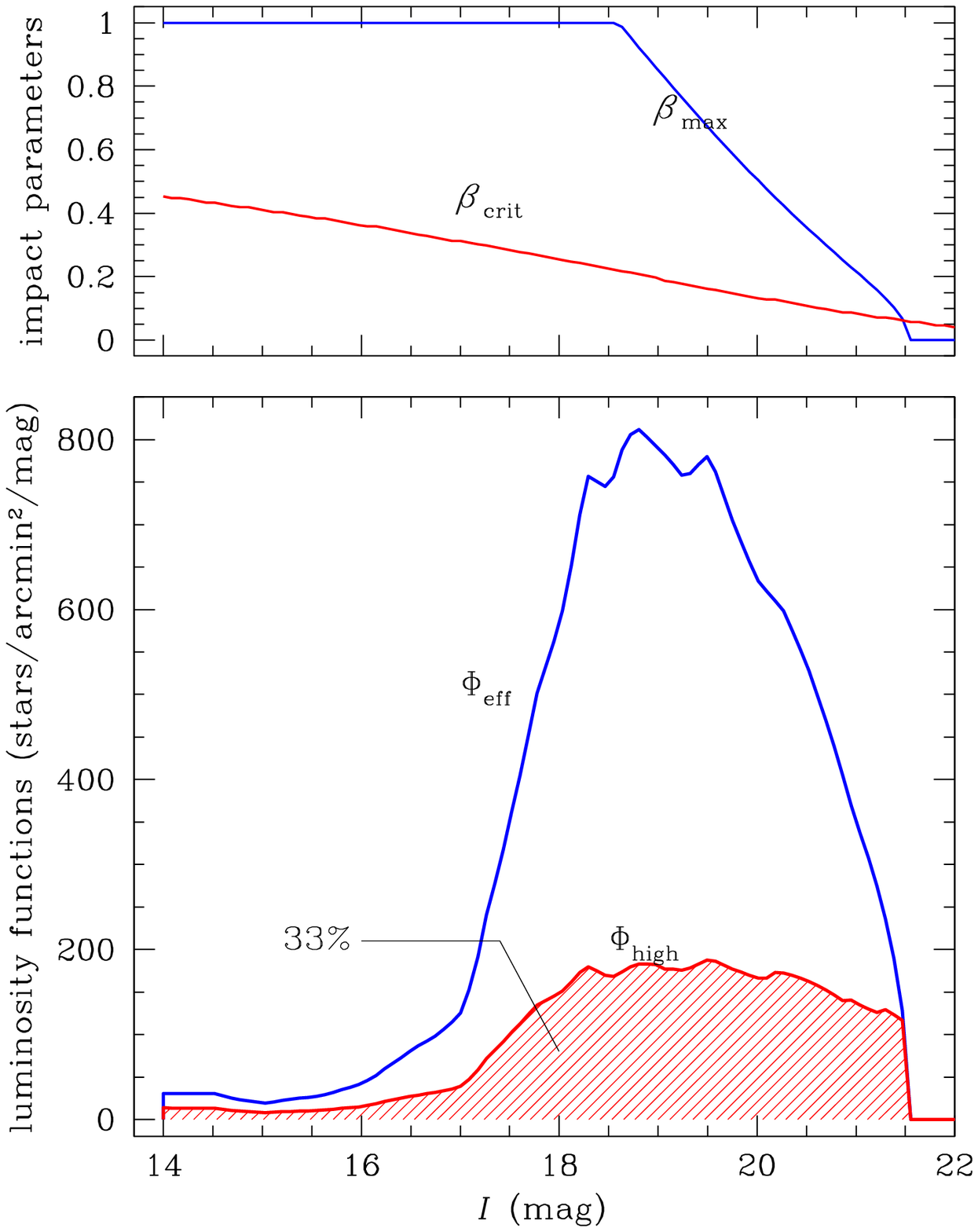}{0.95}
\noindent
{\footnotesize {\bf Figure 4:}\
Upper panel: 
the critical and the maximum impact parameters ($\beta_{\rm crit}$ and 
$\beta_{\rm max}$) as functions of the source brightness for stars in the 
Galactic bulge field. The value of $\beta_{\rm max}$ is equivalent to the 
average detection probability of an event with a source star brightness $I$ 
from the microlensing search by using the DIA method. On the other hand, 
$\beta_{\rm crit}$ represents the average probability that the baseline flux 
of an event with a source star brightness $I$ can be determined with an 
uncertainty less than 50\%.
Lower panel: 
the effective source star luminosity functions of the total ($\Phi_{\rm eff}$) 
and high amplification events ($\Phi_{\rm high}$) detectable from the 
microlensing searches by using the DIA technique toward the Galactic bulge 
field.  The shaded region represents the fraction of high amplification events.
}\clearpage

\end{document}